\pdfoutput=1

\documentclass{article}
\usepackage{amsmath}
\usepackage{graphicx}
\usepackage{color}
\usepackage{amsfonts}
\usepackage{amssymb}
\def\phi{\varphi}

\begin{document}

\title{Teaching the third law of thermodynamics\footnote{\tiny The Open Thermodynamics Journal, 2012, 6, 1-14; \copyright by the author; this article can be distributed freely for academic purposes}}
\author{A. Y. Klimenko
\\The University of Queensland, SoMME, QLD 4072, Australia}
\maketitle
\begin{abstract}
This work gives a brief summary of major formulations of the third law of
thermodynamics and their implications, including the impossibility of
perpetual motion of the third kind. The last sections of this work review more
advanced applications of the third law to systems with negative temperatures
and negative heat capacities. The relevance of the third law to protecting the
arrow of time in general relativity is also discussed. Additional information,
which may useful in analysis of the third law, is given in the Appendices.

This short review is written to assist lecturers in selecting a strategy for
teaching the third law of thermodynamics to engineering and science students.
The paper provides a good summary of the various issues associated with the
third law, which are typically scattered over numerous research publications
and not discussed in standard textbooks.
\end{abstract}

\textbf{Keywords:} laws of thermodynamics, perpetual motion of the third kind,
thermodynamics of black holes; engineering education.

\bigskip

\section{Introduction}

The author of this article has taught senior (year 3) engineering
thermodynamics for a number of years. Engineering students, as probably
everyone else, usually do not have any difficulties in understanding the
concept of energy and the first law of thermodynamics but have more problems
with the concept of entropy, the second and the third laws. This trend in
learning the laws of thermodynamics seems to be quite common. Human
perceptions are well-aligned with the concept of energy but the concept of
entropy tends to be more elusive and typically remains outside the boundaries
of the students' intuition. In thermodynamic courses taught to future
engineers, the concept of entropy is traditionally introduced on the basis of
the Clausius inequality, which directs how to use entropy but does not
explicitly explain the physical nature of the concept. The common perception
that engineering students are incompatible with statistical physics prohibits
the use of the explicit definition of entropy by the Boltzmann-Planck
equation
\begin{equation}
S=k_{B}\ln(\Gamma) \label{BP}%
\end{equation}
where $k_{B}$ is the Boltzmann constant and $\Gamma$ is the number of
micro-states realizing a given macro-state. This equation links the mysterious
entropy $S$ to the fundamental concept of probability, which in this case is
related to $\Gamma$. Despite not being trivial, the concept of probability
combines well with human intuition. My experience is that students very much
appreciate the explicit definition of entropy. The main trend of entropy to
increase, due to the overwhelming probability of occupying whenever possible
macro-states with largest $\Gamma,$ explains the second law.

While statistical physics may offer some insights and assistance in teaching
the second law of thermodynamics, a similar strategy is not likely to work as
an educational remedy for the third law. It seems that there are two major
factors that make the third law and its implications more difficult to
understand than the second law.

First, after more than 100 years since Walter Nernst published his seminal
work \cite{Nernst1906}, which became the beginning of the third law of
thermodynamics, we still do not have a satisfactory universal formulation of
this law. Some formulations of the law seem to be insufficiently general while
others cannot avoid unresolved problems. Many different formulations of the
third law are known; some of them differ only in semantics but many display
significantly different physical understandings of the law.

Another problem in teaching the third law is its abstract character, which,
when compared with the other laws, is apparently less related to core
engineering concepts such as thermodynamic cycles and engines. In many
textbooks the third law is presented as a convenient approach for generating
thermodynamic tables by using absolute entropy. Some formulations of the third
law may not have clear physical implications. Finding a concise and
transparent summary of the implications of the third law of thermodynamics is
still not easy.

This work is an attempt to fill this gap and assist in selecting an approach
to teaching the third law. First we discuss major alternative formulations of
the third law. Then, by analogy with the zeroth, first and second laws, which
prohibit the perpetual motion of the zeroth, first and second kind, we
logically extend this sequence and declare that the third law prohibits the
perpetual motion of the third kind, which can deliver 100\% conversion of heat
into work. This interpretation pertains to the original works of Nernst
\cite{Nernst1912} and is reflected in more recent publications \cite{Som1956}.
One should be aware that the term ``perpetual motion of the third kind'' is
sometimes used to denote a completely different process --- a motion without
friction and loses.

The second law, the third law and the rest of thermodynamics can also be
introduced on the basis of the adiabatic accessibility principle originated by
Constantine Caratheodory \cite{Cara1909}. While this possibility has been
convincingly proven by Lieb and Yngvason \cite{EntOrd2003}, their approach is
rigorous but rather formal and not suitable for teaching. The recent book by
Thess \cite{Thess2011} fills the existing gap and presents adiabatic
accessibility in a very interesting and even entertaining manner. The author
has not had the opportunity to try this approach in class but believes that
adiabatic accessibility can be appreciated by the students and might
eventually become the mainstream approach to teaching thermodynamics.

The last part of this paper is dedicated to some less common but still very
interesting topics in thermodynamics -- the third law in conditions of
negative temperatures or negative heat capacities. The profound connection of
the third law with persevering causality in general relativity seems to be
especially significant. These more advanced topics can be used to stimulate
the interest and imagination of the top students but, probably, would not be
suitable for the rest of the class.

\section{Statements of the third law}

While it was Walter Nernst whose ingenious intuition led thermodynamics to
establishing its third law \cite{Nernst-Nobel}, this important scientific
endeavour was also contributed to by other distinguished people, most notably
by Max Planck and Albert Einstein. The third law of thermodynamics has evolved
from the Nernst theorem -- the analysis of an entropy change in a reacting
system at temperatures approaching absolute zero --, which was first proposed
by Nernst and followed by a discussion between him, Einstein and Planck. Even
after 100 years since this discussion took place, there is still no
satisfactory universal formulation of the third law thermodynamics. Leaving
historical details of this discussion aside (these can be found in an
excellent paper by Kox \cite{Kox2005}), we consider major formulations of the
third law. In this consideration, we follow the broad ideas expressed by the
founders of the third law rather than their exact words --- on many occasions
clear statements of the third law were produced much later
\cite{Simon1937,FG1949}. For example, Nernst did not like entropy, which is
now conventionally used in various statements of the third law, and preferred
to express his analysis in terms of availability. The existing statements
differ not only by semantics but also have significant variations of the
substance of the law; although statements tend to be derived from the ideas
expressed by Planck, Nernst or Einstein and can be classified accordingly.

The most common formulation of the third law of thermodynamics belongs to Max
Planck \cite{Planck1911} who stated that

\begin{itemize}
\item \textbf{Planck formulation.} \textit{When temperature falls to absolute
zero, the entropy of any pure crystalline substance tends to a universal
constant (which can be taken to be zero)}
\begin{equation}
S\rightarrow0\text{ \ \ as \ \ }T\rightarrow0 \label{P3}%
\end{equation}
\end{itemize}

Entropy selected according to $S=0$ at $T=0$ is called absolute. If $S$
depends on $x$ (where $x$ may represent any independent thermodynamic
parameter such as volume or extent of a chemical reaction), then $x$ is
presumed to remain finite in (\ref{P3}). The Planck formulation unifies other
formulations given below into a single statement but has a qualifier ``pure
crystalline substance'', which confines application of the law to specific
substances. This is not consistent with understanding the laws of
thermodynamics as being the most fundamental and universally applicable
principles of nature. This formulation does not comment on entropy of other
substances at $T=0$ and thus is not universally applicable.

The Planck formulation, in fact, necessitates validity of two statements of
unequal universality: the Einstein statement and the Nernst theorem.

\begin{itemize}
\item \textbf{Einstein statement.} \textit{As the temperature falls to
absolute zero, the entropy of any substance remains finite }
\begin{equation}
S(T,x)\rightarrow S_{0}(x),\text{ }\left|  S_{0}\right|  <\infty\text{\ \ as
\ \ }T\rightarrow0,\;\left|  x\right|  <\infty\label{E3}%
\end{equation}
\end{itemize}

The limiting value $S_{0}$ may depend on $x,$ which is presumed to remain
finite at $T\rightarrow0$. Considering expression for the entropy change in a
constant volume heating process
\begin{equation}
S=\int_{0}^{T}\frac{C_{V}}{T}dT \label{S-Cv}%
\end{equation}
it is easy to see that (\ref{E3}) presumes vanishing heat capacity
\begin{equation}
C_{V}\rightarrow0\text{ \ \ at \ \ }T\rightarrow0 \label{E3Cv}%
\end{equation}
since otherwise, the integral in (\ref{S-Cv}) diverges and $S\rightarrow
-\infty$ as $T\rightarrow0$. A similar conclusion can be drawn for $C_{P}$ by
considering the heating process with constant pressure.

The statement is attributed to Einstein \cite{Einstein1907}, who was first to
investigate entropy of quantum systems at low temperatures and to find that
the heat capacities should vanish at absolute zero; this implies that $S $ is
finite at $T\rightarrow0$. Nernst and his group at University of Berlin
undertook extensive experimental investigation of physical properties at low
temperatures, which confirmed the Einstein statement \cite{Kox2005}. It should
be noted that thermodynamic systems become quantized at low temperatures and
classical statistics is likely to produce incorrect results, not consistent
with the Einstein statement. Hence experimental confirmation of the Einstein
statement was at the same time a confirmation for quantum mechanics.\ The
quantum theory of heat capacity was latter corrected by Debye \cite{Debye1912}
to produce a better quantitative match with Nernst's experimental results.
This correction, however, does not affect the validity of the Einstein
statement. Although the validity of the Einstein statement is beyond doubt,
this statement does not capture all important thermodynamic properties at the
limit $T\rightarrow0$.

\begin{itemize}
\item \textbf{Nernst (heat) theorem}. \textit{The entropy change of a system
in any reversible isothermal process tends to zero as the temperature of the
process tends to absolute zero.}
\begin{equation}
S(T,x)-S(T,x+\Delta x)\rightarrow0\text{ \ \ as \ \ }T\rightarrow0,\;\left|
x\right|  <\infty,\;\left|  \Delta x\right|  <\infty\label{NT3}%
\end{equation}
\end{itemize}

The change in $x$ is presumed to remain finite at $T\rightarrow0$. Assuming
``smooth'' differentiation, the Nernst theorem obviously implies that
\begin{equation}
\left(  \frac{\partial S}{\partial x}\right)  _{T}\rightarrow0\text{ as
\ \ }T\rightarrow0 \label{NT3a}%
\end{equation}

The Nernst theorem, although valid in many cases, is unlikely to be universal.
On many occasions Einstein disputed Nernst's arguments aimed at deriving the
heat theorem from the second law \cite{Kox2005}. The main problem with the
theorem is that the entropy $S$ cannot be independent of $x$ at $T=0$ when
some uncertainties are allowed to remain in the system at $T=0$. Indeed, if a
mixture of two or more components (which can be different substances or
different isotopes of the same substance) can be brought to the state of
$T=0,$ then there must be uncertainties in positions of the molecules
representing specific components of the mixture, since the positions of two
different molecules can be swapped to form a new microstate. Assuming that $x$
is the molar fraction of one of the components, we conclude that presence of
these uncertainties should depend on $x$. The Planck formulation unifies two
independent statements -- the Einstein statement and the Nernst theorem -- and
patches the Nernst theorem by restricting application of the Planck
formulation to pure crystalline substances. The third law represents a
statement which physically is related to the second law, but logically is
independent from the second law.

Another formulation of the third law is represented by the following principle:

\begin{itemize}
\item \textbf{Nernst (unattainability) principle}. \textit{Any thermodynamic
process cannot reach the temperature of absolute zero by a finite number of
steps and within a finite time.}
\end{itemize}

The Nernst principle was introduced by Nernst \cite{Nernst1912} to support the
Nernst theorem \cite{Nernst1906} and to counter Einstein's objections. The
Nernst principle implies that an isentropic process (adiabatic expansion or a
similar reversible adiabatic process that can be used to reduce temperature
below that of the environment) cannot start at any small positive $T$ and
finish at absolute zero when volume and other extensive parameters remain
limited, that is
\begin{equation}
S(T,x)-S(0,x+\Delta x)>0\text{ \ \ when \ \ }T>0,\;\left|  x\right|
<\infty,\;\left|  \Delta x\right|  <\infty\label{NS3}%
\end{equation}
If expression (\ref{NS3}) is not valid and $S(T,x)=S(0,x+\Delta x),$ then the
isentropic process starting at $(T,x)$ and finishing at $(0,x+\Delta x)$
reaches the absolute zero. According to the present understanding of the
Nernst principle, $S(T,x)=S(0,x+\Delta x)$ might be possible but then the
isentropic process connecting $(T,x)$ and $(0,x+\Delta x)$ must be impeded by
other physical factors, for example, the process may require an infinite time.

Equivalence of the Nernst principle and the Nernst theorem has repeatedly been
proven in the literature \cite{Gug1967}. These proofs are illustrated by
Figure 1(a,b) demonstrating the possibility or impossibility of an isentropic
process reaching $T=0$ while $x_{1}\leq x\leq x_{2}$ (the bounding lines
represent $x=x_{1}$ and $x=x_{2}$). Case (a) corresponds to the validity of
the Nernst principle and the Nernst theorem while case (b) violates both of
these statements. Achieving $T=0$ in case (a) requires an infinite number of
steps as shown in the figure. The Carnot cycle reaching $T=0,$ which is called
the Carnot-Nernst cycle, is possible in case (b) and is also shown in the
figure. It should be noted that proofs of equivalence of Nernst principle and
Nernst theorem involve a number of additional assumptions \cite{Lan1956,
WHEELER1991} as illustrated by examples (c), (d), (e) and (f). Cases (c) and
(e) indicate that the Nernst principle can be valid while the Nernst theorem
is not. Case (c) violates the Einstein statement while case (e) allows for
fragmented dependence of $S$ on $x$ and shows boundaries $x_{1}\leq x\leq
x_{2}$ and $x_{3}\leq x\leq x_{4}$. Cases (d) and (f) demonstrate validity of
the Nernst theorem and violation of the Nernst principle. Case (d) considers a
special entropy state with $S(T,x)=0$ at $T>0$ - a system with these
properties and Bose-Einstein statistics is discussed by Wheeler
\cite{WHEELER1991}. Case (f) implies negative heat capacities due to $\partial
S/\partial T<0$ at $T\rightarrow0$ and $x=x_{1}$.

While universality of the Nernst theorem is doubtful, it seems that the Nernst
principle has better chances of success. The difficulty of reaching $T=0$ is
supported by experimental evidence. Interactions of a paramagnetic with a
magnetic field are commonly used to reach low absolute temperatures. The
lowest recorded experimental temperature of $T=10^{-10}$K was achieved in a
piece of rhodium metal by YKI research group at Helsinki University of
Technology in 2010 \cite{LTL2010} (this report needs further confirmation).
Unattainability of $T=0$ can be explained by limitations imposed by the Nernst
theorem, when this theorem is valid, or by other restrictions, when the Nernst
theorem is incorrect. For example, reversible transition between mixed and
unmixed states requires selectively permeable membranes; diffusion between
components and through these membranes is likely be terminated at $T=0$. Thus,
although the process illustrated at Figure 1(b) is possible for mixtures, this
process may need an infinite time to complete. We use a weakened version of
the Nernst principle referring to both ``finite number of steps'' and ``finite
time'' to stress its non-equivalence with the Nernst theorem.

Wreszinski and Abdalla \cite{AA-3Law} recently gave new formulation of the
third law, which is based on the concept of adiabatic accessibility
\cite{Cara1909,EntOrd2003,Thess2011}, stating that zero temperatures $T=0$ are
adiabatically inaccessible from any point where $T>0$. The formulation is
equivalent to the Nernst principle. This work \cite{AA-3Law} also includes a
proof of the Nernst theorem from the Nernst principle using the adiabatic
accessibility concept and imposing two additional conditions, which exclude
cases shown in Figures 1(c) and 1(e).

In a summary, we have two major formulations of the third law: the most
reliable but relatively weak Einstein statement and somewhat less certain but
more informative Nernst principle. Currently available scientific evidence
tends to support the validity of the Nernst principle. The Nernst theorem,
although linked to the Nernst principle, is not fully equivalent to this
principle. This theorem does not seem to be universal and, as was first noted
by Einstein, is likely to be incorrect as a general statement. The Nernst
theorem, however, should be correct for pure substances, providing useful
information for analysis of thermodynamic properties at $T\rightarrow0$; some
of these properties are given in Appendix \ref{A1}. Planck's formulation
unifies the Nernst theorem with the Einstein statement but is thus applicable
only to pure substances.

It is interesting that the Nernst principle and the Einstein statement can be
combined to produce the following formulation of the third law:

\begin{itemize}
\item \textbf{Nernst (Nernst - Einstein) formulation}. \textit{A thermodynamic
state with zero absolute temperature can not be reached from any thermodynamic
state with a positive absolute temperature through a finite isentropic process
limited in time and space, although the entropy change between these states is
finite. }
\end{itemize}

This statement implies that
\begin{equation}
0<S(T,x)-S(0,x+\Delta x)<\infty\text{ \ \ when \ \ }0<T<\infty,\;\left|
x\right|  <\infty,\text{ }\left|  \Delta x\right|  <\infty
\end{equation}
or, possibly in some cases, $S(T,x)=S(0,x+\Delta x)$ but the isentropic
process connecting these states needs an infinite time for its completion.
When Nernst \cite{Nernst1912} introduced his unattainability principle, he
formulated and understood this principle in context of validity of the
Einstein statement. If the Einstein statement is not valid and $S\rightarrow
-\infty$ as $T\rightarrow0$, \ then unattainability of $T=0$ is quite obvious.

\section{Perpetual motion of the third kind}

The importance of the laws of thermodynamics is not solely related to their
formal validity; these laws should have a clear physical meaning and applied
significance. Thermodynamics has a very strong engineering element embedded
into this discipline. We might still not know whether irreversibility of the
real world is related only to the temporal boundary conditions imposed on the
Universe or there is some other ongoing fundamental irreversibility weaved
into the matter. Conventional physics, including both classical and quantum
mechanics (but not the interactions of quantum and classical worlds, which may
cause quantum decoherence), is fundamentally reversible. Thermodynamics may
not have all of the needed scientific explanations, but it admits the obvious
(i.e. the irreversibility of the surrounding world), postulates this in form
of its laws and proceeds further to investigate their implications.

The fundamental implications of the laws of thermodynamics are related to
engines - devices that are capable of converting heat into work. Converting
work into heat is irreversible: all work can be converted to heat but not all
heat can be converted to work. The zeroth, first and second laws of
thermodynamics impose restrictions that prohibit certain types of engines ---
these can be conventionally called perpetual motions of the zeroth, first and
second kind depending on which laws these engines violate. The n-th law of
thermodynamics can be formulated by simply stating that perpetual motion of
the n-th kind is impossible. These perpetual engines are illustrated in Figure
2. The first engine represents a possible engine placed into an impossible
situation banned by zeroth law of thermodynamics, when temperatures of the
reservoirs are not transitive, which symbolically can be represented by
$T_{H}>T_{C}>T_{0}>T_{H},$ where $T_{H}>T_{C}$ means that the heat naturally
flows from $T_{H}$ to $T_{C}$. Note that the second law is also violated by
this setup. The second engine illustrates the impossibility of producing work
out of nothing, which is banned by the first law. The third engine produces
work out of heat without any side effects --- this violates the second law.

As in case of the other laws, the third law should have a clear physical
interpretation. Since we have two formulations -- the Einstein statement and
the Nernst principle - we consider two corresponding versions of perpetual
motion of the third kind.

The first perpetual engine of the third kind (Figure 3(a)) violates the
Einstein statement of the third law: it uses the Carnot-Nernst cycle with a
compact cooling reservoir at $T_{C}=0$ and infinitely small entropy
$S=-\infty$ (the cycle's working fluid is presumed to have vanishing heat
capacity at $T\rightarrow0$). The amount of heat disposed by the Carnot cycle
into the cooler is zero under these conditions ($Q_{C}=Q_{H}T_{C}%
/T_{H}\rightarrow0$ as $T_{C}\rightarrow0$) and all of $Q_{H}$ is converted
into work. The cooler, however, must receive the entropy $\Delta S_{H}$ lost
by the heater. Since its entropy is infinitely negative, the state of the
cooler is not affected by this entropy dump. One can unify the Carnot-Nernst
cycle with the compact cooler and call this an engine converting heat into
work. This perpetual engine clearly contradicts if not the letter then the
spirit of the second law. Unlimited entropy sinks, which allow for extraction
of unlimited work from the environment, are banned by the Einstein statement
of the third law. The physical meaning of this statement is a thermodynamic
the declaration of existence of quantum mechanics, which does not allow
allocation of unlimited information within a limited volume due to quantum uncertainty.

The second perpetual engine of the third kind shown in Figure 3(b) violates
the Nernst principle and works with a large cooling reservoir at $T=0$, which
serves as an entropy sink. The engine uses the Carnot-Nernst cycle and
converts 100\% of heat into work while dumping the excess entropy $\Delta
S_{H}$ into the cooling reservoir. The Nernst principle prohibits reaching
$T=0$ in the cycle and does not allow conversion of 100\% of heat into work
under these conditions. Entropy can be interpreted as negative information -
i.e. absence of information about the exact micro-state of the system. The
Nernst principle allows for reduction of the energy content of information by
lowering $T$ but does not permit the complete decoupling of information and
energy that occurs at $T=0$.

\section{Negative temperatures}

Thermodynamic systems may have negative temperatures \cite{Ramsey1956,LL5}. A
simple thermodynamic system involving only two energy levels is sufficient to
bring negative temperatures into consideration. The thermodynamic relations
for this system are derived in Appendix \ref{A2}. The entropy $S$ and inverse
temperature $1/T,$ obtained in Appendix \ref{A2},\ are plotted in Figure 4
against energy $E$. It can be seen that the region of negative temperatures
lies above the region of positive temperatures with $T=+0$ being the lowest
possible temperature and $T=-0$ being the highest possible temperature. As the
energy of the system increases from $E=0$, particles may now be allocated at
both energy levels and this increases uncertainty and entropy. As the energy
increases further towards it maximal value $E=E_{1},$ most of the particles
are pushed towards the high energy level and this decreases uncertainty and
entropy. Note that the function $S(E)$ is symmetric and $T(E)$ is
antisymmetric with respect to the point $E=E_{1}/2$ for this example.

We can define quality of energy as being determined by the function
$\beta=-1/T$ \ so that higher quality corresponds to larger (more positive)
$\beta$. The quality of work corresponds to the quality of heat at $T=\infty$
and $\beta=0$. Energy can easily lose its quality and be transferred from
higher to lower $\beta$ but upgrading the quality of energy is subject to the
usual restrictions of the second law of thermodynamics. This law can be
formulated by stating that energy cannot be transferred from a lower quality
state to a higher quality state without any side effects on the environment.

The state of negative temperatures is highly unstable and it cannot have
conventional volume as separating the volume into smaller parts moving
stochastically with macro-velocities increases the entropy. Indeed, the
quality of kinetic energy (work) is $T=\infty$ and is below the quality of
energy at negative temperatures. For these systems, shattering into small
pieces appears to be thermodynamically beneficial. A strict proof of this
statement is given in the famous course on theoretical physics of Landau \&
Lifshits\ \cite{LL5}. Negative temperatures may, however, exist in systems
that do not posses macroscopic momentum. Both Ramsey \cite{Ramsey1956} and
Landau \& Lifshits \cite{LL5} nominate a physical system that may possess
negative temperatures during time interval of a measurable duration:
interactions of nuclear spins with each other and a magnetic field in a
crystal. After a fast change in direction of magnetic field this system is in
a state with a negative temperature, which will last until the energy is
transferred to the rest of the\ crystal.

Now we return to the third law of thermodynamics. It is most likely that
difficulties and restrictions for reaching positive zero $T=+0,$ which are
stated by the third law of thermodynamics, are also applicable to negative
zero $T=-0$. \ In the absence of experimental data, this statement may seem
speculative but, considering the difficulties of reaching any negative
temperatures, achieving $T=-0$ would not be any easier than achieving $T=+0$.
In the example given in Appendix \ref{A2}, an instantaneous reverse of the
direction of the magnetic field changes state $E$ to $E_{0}-E$ so that $T$ is
changed to $-T$ . Hence reaching $T=-0$ allows us to reach $T=+0$ and vice versa.

\section{Negative heat capacities}

Thermodynamic systems with negative heat capacities $C<0$ are unusual objects
\cite{Bell1999}. In particular, they cannot be divided into equilibrated
subsystems, say A and B, and, as proven by Schr\"{o}dinger \cite{Shr1952} (see
Appendix \ref{A3}), and cannot be treated by conventional methods of
statistical physics (i.e. using the partition function) since these methods
imply the existence of equilibrium between subsystems. Indeed, if A and B are
initially at equilibrium and $C_{\text{A}}<0$ and $C_{\text{B}}<0$ this
equilibrium is unstable. Let a small amount of heat $\delta Q$ to be passed
from A to B, then $T_{\text{A}}$ tend to increase and $T_{\text{B}} $ tend to
decrease and this encourages further heat transfer from A to B which further
increases $T_{\text{A}}$ and decreases $T_{\text{B}}$. The initial equilibrium
between A and B is unstable.

An object with $C<0$ can however be in equilibrium with a reservoir having
positive capacity $C_{r}>0$ provided $\left|  C\right|  >C_{r}$. Indeed, if
initially $T=T_{r}$ and $\delta Q$ is passed from the object to the reservoir;
both $T$ and $T_{r}$ increase but according to condition $\left|  C\right|
>C_{r},$ the temperature $T_{r}$ increases more than $T,$ encouraging heat
transfer back from the reservoir to the object.

Although thermodynamic states with negative heat capacity are unstable, such
cases have been found among conventional thermodynamic objects with a short
existence time \cite{sodium147}. Here, we consider thermodynamic objects with
persistent negative heat capacities $C<0,$ and term them thermodynamics stars
and thermodynamic black holes due to the vague similarity of their
thermodynamic properties to those of real stars and black holes. The heat
capacity of a star is negative in Newtonian gravity as considered in Appendix
\ref{A3a}.

\subsection{Thermodynamic stars and black holes}

The environment is a reservoir with a very large size and capacity so its
temperature $T_{0}$ does not change. Equilibrium of an object with negative
$C$ and the environment is always unstable. Assume that the temperature of the
object $T$ is slightly above that of the environment $T_{0}$. In this case the
object tends to lose some energy due to heat transfer. Since its heat capacity
is negative, this would further increase $T$ resulting in more energy loss.
The process will continue until the object loses all of its energy; due to
obvious similarity we will call these cases \textit{thermodynamic stars}. If a
thermodynamic star loses its energy at a rapidly increasing pace as determined
by its rapidly rising temperature, it may explode --- i.e. reach negative
temperatures before losing all of its energy and then disintegrate into small
pieces as discussed in the previous section.

We, however, are more interested in an opposite case when $T$ is slightly
below the environmental temperature $T_{0}$. In this case energy tends to be
transferred to the object from the environment, resulting in further
temperature $T$ decrease and energy$\ E$ increase. As $T$ drops to very low
values, extracting energy from the object by thermodynamic means becomes
practically impossible (as this would need even smaller temperatures). In this
case the object can be termed a \textit{thermodynamic black hole}. Depending
on the nature of the limiting state $T\rightarrow0$ we divide thermodynamic
black holes into three types:

\begin{itemize}
\item \textbf{Type 1:} $\ S$ and $E$ remain bounded. This type is consistent
with the Einstein statement but may violate the Nernst theorem and the Nernst principle.

\item \textbf{Type 2:} $E$ remains bounded but $S$ is not. This type obviously
violates the Einstein statement and most likely the Nernst theorem.

\item \textbf{Type 3:} both $S$ and $E$ are not bounded. This type complies
with the Nernst principle but may violate the other statements.
\end{itemize}

These types of thermodynamic black holes are illustrated in Figure 5. Note
that, according to equation (\ref{CSE}), $\partial^{2}S/\partial E^{2}$ is
positive when $C$ is negative (hence the case $S\rightarrow
\operatorname{const},$ $E\rightarrow\infty$ cannot occur). A type 1 black hole
reaching $T=0$ cannot lose any heat since it has the lowest possible
temperature and cannot gain any heat since its gaining capacity is saturated
-- it is thermodynamically locked from its surroundings.

We now examine gravitational black holes, whose major characteristics are
listed in Appendix \ref{A4}.

\subsection{Schwarzschild black holes}

The Schwarzschild black holes are the simplest type of gravitational black
holes and are controlled by a single parameter -- the mass of the hole $M$ --,
which determines the radius, surface area and volume of the hole
\cite{Davies1978}
\begin{equation}
r_{S}=2\frac{\gamma}{c^{2}}M,\;\;A=4\pi r_{S}^{2},\;\;V=\frac{4}{3}\pi
r_{S}^{3}%
\end{equation}
Here we refer, of course, to the dimensions of the event horizon, which, for
Schwarzschild black holes, is a sphere surrounding the time/space singularity.
Nothing, not even light, can escape from within the event horizon. It is
interesting that the relativistic expression for the radius of the
Schwarzschild event horizon $r_{S}$ coincides with the corresponding Newtonian expression.

The Einstein energy, Bekenstein-Hawking entropy and Hawking temperature of the
black hole are given by \cite{Davies1978}
\begin{equation}
E=Mc^{2},\;\;S=4\pi\frac{\gamma k_{B}}{c\hbar}M^{2},\;\;T=\frac{\hbar c^{3}%
}{8\pi\gamma k_{B}}\frac{1}{M}%
\end{equation}
These equations are combined into conventional
\begin{equation}
dE=TdS \label{dETdS}%
\end{equation}
Note that Schwarzschild black holes have negative heat capacities
\begin{equation}
C=T\frac{\partial S}{\partial T}=\frac{\partial S}{\partial M}\left(
\frac{\partial\ln(T)}{\partial M}\right)  ^{-1}=-2S
\end{equation}
The Bekenstein entropy can be estimated from the quantum uncertainty principle
$\Delta E\;\Delta t\sim\hbar$ where $\Delta E$ is minimal energy of a quantum
wave and $\Delta t$ is its maximal life time. Inside the horizon, the radial
coordinate $r$ becomes time-like (one can say that time $t$ and\ space
distance $r$ ``swap'' their coordinates) and $\Delta t\sim r_{S}/c,$ hence
$\Delta E\sim\hbar c^{3}/(\gamma M).$ The ratio $E/\Delta E$\ then represents
an estimate for the maximal number of quantum waves within the horizon.
Assuming that each wave may have at least $2$ states, say with positive and
negative spins, we obtain the following estimate for the corresponding number
of macro-states\ $\Gamma\sim2^{E/\Delta E}\sim\exp(E/\Delta E).$\ The
Boltzmann-Planck equation (\ref{BP}) indicates that\ $S\sim k_{B}E/\Delta
E\sim k_{B}\gamma M^{2}/(\hbar c)$. This estimate and equation (\ref{dETdS})
necessitate that $T\sim\hbar c^{3}/(\gamma k_{B}M)$. According to a more
rigorous theory developed by Hawking \cite{Hawking1974} (in the wake of
Zeldovich's \cite{Zeld1971,Kip1994} analysis indicating that rotating black
holes emit radiation), black holes can radiate due to quantum fluctuations
appearing everywhere including the event horizon. As field disturbances
propagate away from the hole, they experience red shift due to relativistic
time delays in strong gravity. The Hawking temperature is the effective
temperature of a black hole as observed from a remote location. It is useful
to note numerical values of the constants:
\[
T\approx\frac{1.2\times10^{23}\text{kg}}{M}\text{K,\ \ }S\approx
3.6\times10^{-7}M^{2}\frac{\text{J}}{\text{kg}^{2}\text{K}},\;\;r_{S}%
\approx1.48\times10^{-27}M\frac{\text{m}}{\text{kg}}%
\]
It is generally believed that, due to restrictions of quantum mechanics, the
Bekenstein-Hawking entropy of Schwarzschild black holes represents the maximum
possible entropy allocated within a given volume. A black hole works as the
ultimate shredding machine: all information entering the black hole is
destroyed introducing maximal uncertainty (although mass, charge and angular
momentum are preserved). The carrying mass is packed to maximal possible
density and reaches maximal entropy. While in absence of gravity the state of
maximal entropy is achieved by a uniform dispersal of a given matter over the
available volume, shrinking the same matter into a singularity point is
favored by the second law of thermodynamics in the presence of a gravitational
pull. The relatively small number of black holes, that presumably existed in
the early Universe, is responsible for its initial low-entropy state that
provides thermodynamic exergy needed for powering stars and galaxies. The
Universe works as if it was a very large thermodynamic engine!

We now examine compliance with the formulations of the third law. In terms of
the classification given above, Schwarzschild black holes are of type 3 and
they clearly comply with the Nernst principle. Since mass and volume are not
restricted at the limit $T\rightarrow0$ (and so is the specific volume
$V/M\rightarrow\infty$), the Nernst theorem is not formally violated due to
non-compliance with its condition of keeping the secondary thermodynamic
parameter $x$ finite. The thermodynamic quantities characterizing the black
hole behave differently as $T\rightarrow0$ depending on whether they are
considered on ``per mass'' or ``per volume'' basis. Entropy and capacity per
volume comply with the Planck and Einstein statements and the Nernst theorem
$C/V\rightarrow-0,$ $S/V\rightarrow+0$ as $T\rightarrow0$ and $M\rightarrow
\infty,$ while the same quantities per mass do not: $C/M\rightarrow
-\infty,\;S/M\rightarrow+\infty$ as $T\rightarrow0$ and $M\rightarrow\infty$.

\subsection{Kerr-Newman black holes}

Kerr-Newman black holes are rotating and electrically charged black holes,
characterized by three parameters: mass $M$, angular momentum $J$ and charge
$q$. These black holes have a very complex space-time structure, which
possesses only cylindrical (but not spherical) symmetry. The generalization of
the equations presented in the previous subsection leads us to
\cite{Davies1978}
\begin{equation}
T=\frac{\hbar}{ck_{B}}\frac{\kappa}{2\pi},\;\;S=k_{B}\frac{A}{4l_{p}^{2}%
},\;\;l_{p}^{2}=\frac{\gamma\hbar}{c^{3}}%
\end{equation}%
\begin{equation}
dE=TdS+\Omega dJ+\Phi dq
\end{equation}
where $\kappa$ is the surface gravity and $l_{p}$ is the Planck length scale.
The definitions of the angular rotation speed $\Omega$ and the electrical
potential $\Phi,$ as well as the associated equations expressing $\kappa$ and
$A$ in terms of $M$, $J$ and $q,$ are given in Appendix \ref{A4}. The heat
capacity of Kerr-Newman black hole may be negative or positive depending on
the values of the parameters $M$, $J$ and $q$.

A major feature of Kerr-Newman black holes is\ that zero temperature may be
achieved with finite values of the parameters $M$ $J,$ and $q$; this hole
belongs to type 1. Indeed equation (\ref{SurfGrav}) indicates that $\kappa=0 $
and $T=0$ when
\begin{equation}
\tilde{M}^{2}=\tilde{q}^{2}+\frac{\tilde{J}^{2}}{\tilde{M}^{2}}
\label{extreme}%
\end{equation}
This state of a black hole is called extreme. For the sake of simplicity, we
use normalized (geometric) values of the parameters marked by ``tilde'' and
defined in the Appendix. The entropy in this state tends to a finite limit
\begin{equation}
S\rightarrow k_{B}\frac{A_{0}}{4l_{p}^{2}},\;A\rightarrow A_{0}=4\pi
(2\tilde{M}^{2}-\tilde{q}^{2})\text{ as }T\rightarrow0
\end{equation}
Note that $\tilde{M}^{2}\geq\tilde{q}^{2}$ according to (\ref{extreme}) and
$A_{0}>0$. One can see that this extreme state complies with the Einstein
statement but clearly violates the Nernst theorem. Indeed, the entropy $S$
can, in principle, be changed by dropping suitably selected charged particles
into the black hole and changing $A_{0}$ without altering condition
(\ref{extreme}), that is at $T=0$. If a black hole can physically reach its
extreme state, this would also violate the Nernst principle. This appears to
be quite important for modern physics and is discussed in the rest of this section.

Charged and rotating black holes have not one event horizon but two: the inner
and the outer. As the black hole reaches its extreme state these horizons
approach each other and finally merge. If any further increase in charge and
angular momentum or decrease in mass occurs, and the event horizons disappear
as indicated by $A$ becoming complex in equation (\ref{AMJq}). The
singularity, which is normally hidden behind the horizons, becomes ``naked''.
If this happens anywhere in the Universe, the nature of the Universe changes:
we can gain access to non-chronal regions previously protected by the
horizons, where many wonderful things such as closed time-like curves and time
travel are possible. The implications of this possibility for our
understanding of the Universe involve violations of the causality principle
and are so severe that Roger Penrose suggested the ``cosmic censorship''
principle prohibiting naked singularities \cite{Penrose1973}.

Thermodynamics, which is fundamentally linked to the arrow of time, likes the
possibility of time travel even less than the other sciences. The Nernst
principle, however, prohibits reaching the extreme state since it has $T=0$
and protects causality in general and irreversibility of the second law in
particular. While a rigorous proof of the unattainability statement for
Kerr-Newman black holes can be found in the literature \cite{Israel1986}, we
restrict our consideration to a simple illustration. As any thermodynamic
object, a black hole radiates energy with intensity $dE/dt\sim AT^{4}$. We may
try to reach extreme state by radiating energy (and mass) of the black hole
while keeping its charge the same. If the energy $E_{0}=M_{0}c^{2}$
corresponds to the extreme state, then, according to (\ref{SurfGrav}) and
(\ref{AMJq}), $T\sim\kappa\sim(E-E_{0})^{1/2}$ and $A\rightarrow A_{0}$\ as
$E\rightarrow E_{0}.$ Hence, we obtain $dE/dt\sim(E-E_{0})^{2}$ resulting in
$E-E_{0}\sim1/t$. In accordance with the Nernst principle, an infinite time
(as measured by a remote observer) is needed for a black hole to reach its
extreme state.

\section{Concluding remarks}

The importance of thermodynamic laws lies not only in the formal correctness
of thermodynamic statements but also in their universal applicability and
physical significance. The third law is an independent statement, which acts
as a ``guardian angel'' for the second law. There is still no perfect
formulation for the third law of thermodynamics. Planck's statement and the
Nernst theorem are not universal (or likely to be incorrect if formulated as
general statements). The Einstein statement is, to the best of our knowledge,
correct and universal but it is weaker than generally expected from the third
law of thermodynamics. The considered version of the Nernst principle, which
is not fully equivalent to the Nernst theorem, seems to be both universal and
correct and is a good candidate for the third law. Recent developments in
cosmological thermodynamics tend to support this view: thermodynamics of black
holes violates the Nernst theorem but is considered to uphold the Einstein
statement and the Nernst principle. The relation between the Nernst principle
and protection of causality by cosmic censorship seems profound and produces a
very strong argument in favor of formulating the third law of thermodynamics
on the basis of this principle.

Both the Einstein statement and the Nernst principle have physical and
engineering implications and can be recommended to represent the third law in
thermodynamic courses (separately or as the combined Nernst-Einstein
formulation). It is probably better to keep the existing uncertainty in
choosing the best formulation of the third law outside the scope of
engineering thermodynamic courses --- this would be confusing for the
students. Selecting a single general formulation and focusing on physical
implications rather than on mathematical strictness of the formulation seems
to be the best approach. The engineering interpretation of the law --- no
perpetual motion of the third kind is possible --- is a way to induce the
students' interest and convince them of the importance of the law. The best
engineering and science students may also benefit from a wider discussion of
the role of the laws of thermodynamics in the Universe.

\section{Acknowledgment}

This article was specifically written for Open Journal of Thermodynamics and,
unlike other works of this author, is not supported by any grants.

\pagebreak \bigskip

{\Huge APPENDICES}

\appendix

\section{Thermodynamic relations \label{A1}}

This Appendix presents thermodynamic relations, which can be useful in
analysis of the third law and its implications \cite{LL5}. We begin with the
well-known Maxwell relations which are derived from the commutative properties
of partial derivatives (for example $\partial^{2}G/(\partial T\partial
P)=\partial^{2}G/(\partial P\partial T))$. Differentials of energy $E,$
enthalpy $H,$ Gibbs $G$ and Helmholtz $F$ free energies indicate
\begin{equation}
\frac{\partial(S,T)}{\partial(P,T)}=-\frac{\partial(V,P)}{\partial(T,P)}%
=\frac{\partial G}{\partial T\partial P},\;\;\frac{\partial(S,T)}%
{\partial(V,T)}=\frac{\partial(P,V)}{\partial(T,V)}=\frac{\partial F}{\partial
T\partial V}\label{MX1}%
\end{equation}%
\begin{equation}
\frac{\partial(T,S)}{\partial(V,S)}=-\frac{\partial(P,V)}{\partial(S,V)}%
=\frac{\partial E}{\partial S\partial V},\;\;\frac{\partial(T,S)}%
{\partial(P,S)}=\frac{\partial(V,P)}{\partial(S,P)}=\frac{\partial H}{\partial
S\partial P}\label{MX2}%
\end{equation}
Here partial derivatives are expressed in terms of Jacobians --- this can be
quite useful in derivations. For example $\partial S/\partial T$ with constant
$V$ can be equivalently interpreted as Jacobian for replacement of variables
$(T,V)$ by $(S,V)$. \ The following definitions are related to heat
capacities
\begin{equation}
C_{V}=T\left(  \frac{\partial S}{\partial T}\right)  _{V}=T\frac
{\partial(S,V)}{\partial(T,V)}=\left(  \frac{\partial E}{\partial T}\right)
_{V}%
\end{equation}%
\begin{equation}
C_{P}=T\left(  \frac{\partial S}{\partial T}\right)  _{P}=T\frac
{\partial(S,P)}{\partial(T,P)}=\left(  \frac{\partial H}{\partial T}\right)
_{P}%
\end{equation}%
\begin{equation}
\frac{1}{C_{V}}=\frac{1}{T}\left(  \frac{\partial^{2}E}{\partial S^{2}%
}\right)  _{V}=\frac{-1}{T}\left(  \frac{\partial E}{\partial S}\right)
^{3}\left(  \frac{\partial^{2}S}{\partial E^{2}}\right)  _{V}=-T^{2}\left(
\frac{\partial^{2}S}{\partial E^{2}}\right)  _{V}\label{CSE}%
\end{equation}
The ratio of adiabatic to isothermal compressibilities is linked to heat
capacities
\begin{equation}
\frac{\left(  \frac{\partial P}{\partial V}\right)  _{S}}{\left(
\frac{\partial P}{\partial V}\right)  _{T}}=\frac{\frac{\partial
(P,S)}{\partial(V,S)}}{\frac{\partial(P,T)}{\partial(V,T)}}=\frac
{T\frac{\partial(P,S)}{\partial(P,T)}}{T\frac{\partial(V,S)}{\partial(V,T)}%
}=\frac{C_{P}}{C_{V}}%
\end{equation}
The difference between $C_{P}$ and $C_{V}$ is evaluated by representing
entropy as $S=S(T,V(T,P))$ so that
\[
C_{P}-C_{V}=-T\frac{\partial(S,T)}{\partial(V,T)}\frac{\partial(V,P)}%
{\partial(T,P)}=
\]%
\begin{equation}
=-T\frac{\partial(P,V)}{\partial(T,V)}\frac{\partial(V,P)}{\partial
(T,P)}=T\frac{\partial(S,T)}{\partial(V,T)}\frac{\partial(S,T)}{\partial
(P,T)}\label{CP-CV}%
\end{equation}
Maxwell relations (\ref{MX1}) and (\ref{MX2}) are used here. The Nernst heat
theorem (\ref{NT3}) requires (\ref{NT3a}) that is $(\partial S/\partial
x)_{T}\rightarrow0$ as $T\rightarrow0$ where $x$ can be any thermodynamic
parameter, for example $V$ or $P$. Hence
\begin{equation}
\frac{\partial(S,T)}{\partial(V,T)}=\frac{\partial(P,V)}{\partial
(T,V)}\rightarrow0,\;\;\frac{\partial(S,T)}{\partial(P,T)}=-\frac
{\partial(V,P)}{\partial(T,P)}\rightarrow0\text{ as }T\rightarrow
0\label{Law3-SVP}%
\end{equation}
In conjunction with (\ref{CP-CV}), these relations imply that
\begin{equation}
\frac{C_{P}}{C_{V}}\rightarrow1\text{ as }T\rightarrow0
\end{equation}
The derivative, which indicates the rate of reduction of temperature in
adiabatic expansion,
\begin{equation}
\frac{\partial(T,S)}{\partial(V,S)}=\frac{\partial(T,S)}{\partial(V,T)}%
\frac{\partial(V,T)}{\partial(V,S)}=-\left(  \frac{\partial S}{\partial
V}\right)  _{T}\frac{T}{C_{V}}\rightarrow0\text{ as }T\rightarrow0
\end{equation}
vanishes under restrictions of the Nernst theorem according to (\ref{Law3-SVP}%
). Here we use $V$ and $P=\left(  \partial E/\partial V\right)  _{S}$\ but
these relationships can be generalized for any other consistent thermodynamic
variables, say $x$ and $y=\left(  \partial E/\partial x\right)  _{S}$.

\section{A simple thermodynamic system with negative temperatures \label{A2}}

Let us consider a simple example: $n$ particles can occupy one of the two
energy levels $0$ and $E_{1}$ (this can be practically achieved when, for
example, particle spins can take two values +%
${\frac12}$%
and -%
${\frac12}$%
in a magnetic filed). These levels are degenerate, with $k_{0}$ and $k_{1}$
representing the corresponding degeneracy factors. Classical statistics is
assumed implying that $k_{0},k_{1}\gg n$. There is no need to evaluate the
partition function for this simple case. Indeed, assuming that $n_{1}=n-n_{0}
$ particles are located at the energy level $E_{1}$, the system energy is
determined by
\begin{equation}
E=E_{1}n_{1} \label{NT-E}%
\end{equation}
The $n_{1}$ indistinguishable particles can be allocated on $k_{1}$ sublevels
of the level $E_{1}$ by $\Gamma_{1}=k_{1}^{n_{1}}/n_{1}!$ micro-states.
Similarly, $\Gamma_{0}=k_{0}^{n_{0}}/n_{0}!$ . According to the
Boltzmann-Planck equation (\ref{BP}) entropy $S$ is linked to $\Gamma
=\Gamma_{0}\Gamma_{1}$ by the equation
\begin{equation}
S=k_{B}\ln(\Gamma_{0}\Gamma_{1})=-k_{B}\left(  n_{0}\ln\left(  \frac{n_{0}%
}{ek_{0}}\right)  +n_{1}\ln\left(  \frac{n_{1}}{ek_{1}}\right)  \right)
\label{NT-S}%
\end{equation}
Here we use $\ln(n!)=n\ln(n/e)+...$ and conventionally neglect smaller terms.
With the use of
\begin{equation}
S=-k_{B}n\left(  \left(  1-\frac{E}{E_{1}}\right)  \ln\left(  \frac{n}{ek_{0}%
}\left(  1-\frac{E}{E_{1}}\right)  \right)  +\frac{E}{E_{1}}\ln\left(
\frac{n}{ek_{1}}\frac{E}{E_{1}}\right)  \right)  \label{NT-SE}%
\end{equation}
If $k_{0}=k_{1}=k,$ then
\begin{equation}
S=-k_{B}n\left(  \left(  1-\frac{E}{E_{1}}\right)  \ln\left(  \left(
1-\frac{E}{E_{1}}\right)  \right)  +\frac{E}{E_{1}}\ln\left(  \frac{E}{E_{1}%
}\right)  \right)  +S_{0} \label{NT-SE2}%
\end{equation}
where constant $S_{0}$ is defined\ by
\begin{equation}
S_{0}=-k_{B}n\ln\left(  \frac{n}{ek}\right)  \label{NT-S0}%
\end{equation}
The temperature $T$ is determined by
\begin{equation}
\frac{1}{T}=\frac{dS}{dE}=\frac{k_{B}n}{E_{1}}\ln\left(  \frac{E_{1}}%
{E}-1\right)  =-\frac{k_{B}n}{E_{1}}\ln\left(  \frac{E_{1}}{E_{1}-E}-1\right)
\label{NT-T}%
\end{equation}

\bigskip

\section{Partition function and positiveness of heat capacities \label{A3}}

Schr\"{o}dinger \cite{Shr1952} proved that a system whose thermodynamics is
characterized by the partition function $Z$ and statistical sums
\begin{equation}
\left\langle E\right\rangle =\frac{1}{Z}\sum_{i}E_{i}\exp\left(  -\frac{E_{i}%
}{k_{B}T}\right)
\end{equation}%
\begin{equation}
Z=\sum_{i}\exp\left(  -\frac{E_{i}}{k_{B}T}\right)  ,
\end{equation}
must have a non-negative heat capacity. Indeed
\begin{equation}
C=\frac{\partial\left\langle E\right\rangle }{\partial T}=\frac{1}{Z}\sum
_{i}\frac{E_{i}^{2}}{k_{B}^{2}T^{2}}\exp\left(  -\frac{E_{i}}{k_{B}T}\right)
-\frac{\left\langle E\right\rangle Z}{Z^{2}}\frac{\partial Z}{\partial
T}=\frac{\left\langle E^{2}\right\rangle -\left\langle E\right\rangle ^{2}%
}{k_{B}^{2}T^{2}}\geq0,
\end{equation}
where we use
\begin{equation}
\frac{\partial Z}{\partial T}=\frac{\left\langle E\right\rangle }{k_{B}%
^{2}T^{2}}Z
\end{equation}
This indicates that Gibbs methods based on evaluating the partition function
are not applicable to systems with negative heat capacities.

\section{Newtonian dust star -- a system with negative heat capacity
\label{A3a}}

In astrophysics, the energy of particles interacting by gravity with each
other is
\begin{equation}
E=U+K
\end{equation}
the sum of kinetic energy $K$ and potential energy $U$ of gravitational
interactions. According to the Virial theorem (\ref{Virial}), which is proved
below, $U=-2K$ in classical gravity with the gravitational potential
$\varphi\sim-1/r$. Note that the kinetic energy is always positive, while
potential energy in the gravitational field is negative attaining zero in
remote locations.\ Hence $E=-K$ and the kinetic energy $K$ of $n$ moving
particles is linked to the temperature by \cite{Bell1999}
\begin{equation}
-E=K=\frac{3}{2}nk_{B}T
\end{equation}
Evaluation of heat capacity
\begin{equation}
C=\frac{dE}{dT}=-\frac{3}{2}nk_{B}%
\end{equation}
yields negative values for $C$. Hence, when the particles forming a star lose
energy due to radiation, the temperature of the star increases \cite{Bell1999}.

The Virial theorem is proved by the following equation
\begin{equation}
\underset{\approx0}{\underbrace{\frac{d^{2}}{dt^{2}}\frac{1}{2}\sum_{i}%
m_{i}r_{i}^{2}}}=\underset{K}{\underbrace{\frac{1}{2}\sum_{i}m_{i}\dot{r}%
_{i}^{2}}}+\underset{\frac{a}{2}U}{\underbrace{\sum_{i}m_{i}\mathbf{r}%
_{i}\cdot\mathbf{\ddot{r}}_{i}}} \label{Virial}%
\end{equation}
for a system of $i=1,...,n$ particles with masses $m_{i}$ and locations
$\mathbf{r}_{i}$. In a stationary state, the average distance between
particles does not change and the first term is close to zero. The second term
is obviously linked to kinetic energy $K$ while the connection of the last
term to $U$ can be demonstrated as follows
\[
\sum_{i}m_{i}\mathbf{r}_{i}\cdot\mathbf{\ddot{r}}_{i}=\sum_{i}\sum_{j\neq
i}\mathbf{r}_{i}\cdot\mathbf{F}_{ij}=\frac{1}{2}\sum_{i}\sum_{j\neq i}\left(
\mathbf{r}_{i}-\mathbf{r}_{j}\right)  \cdot\mathbf{F}_{ij}%
\]%
\[
=-\frac{1}{2}\sum_{i}\sum_{j\neq i}r_{ij}\frac{d\varphi_{ij}}{dr_{ij}}%
=\frac{1}{2}\sum_{i}\sum_{j\neq i}a\varphi_{ij}=\frac{a}{2}U
\]
Newton's third law $\mathbf{F}_{ij}=-\mathbf{F}_{ji}=-\nabla\varphi_{ij}$ is
applied here while using potential $\varphi_{ij}(r_{ij})\sim-1/r_{ij}^{a},$
where $\mathbf{r}_{ij}=\mathbf{r}_{i}-\mathbf{r}_{j}$.

\section{Gravitational black holes \label{A4}}

General stationary \textbf{Kerr-Newman black holes} are characterized by three
parameters: mass $M$, angular momentum $J$ and charge $q$. In general
relativity, it is convenient to use so called geometric units, which are
denoted here by the ``tilde'' symbol:
\[
\tilde{M}=M\frac{\gamma}{c^{2}},\;\;\tilde{E}=E\frac{\gamma}{c^{4}}%
,\;\;\tilde{q}=q\frac{1}{c^{2}}\left(  \frac{\gamma}{4\pi\varepsilon_{0}%
}\right)  ^{1/2},
\]%
\begin{equation}
\;\tilde{J}=J\frac{\gamma}{c^{3}},\;\;\tilde{\kappa}=\frac{\kappa}{c^{2}%
},\;\tilde{A}=A
\end{equation}
where $c$ is the speed of light, $\gamma$ is the gravitational constant and
$\varepsilon_{0}$ is the electric constant. Geometric mass $\tilde{M},$ energy
$\tilde{E},$ and charge$\;\tilde{q}$ are measured in length units (m)$;$
geometric angular momentum $\tilde{J}$ and area $\tilde{A}$ are measured in
area units (m$^{2}$); geometric surface gravity $\tilde{\kappa}$ is measured
in inverse length units (1/m). The area of the event horizon -- the surface of
no-return surrounding the black hole -- is given by \cite{Davies1978}%

\begin{equation}
\frac{A}{4\pi}=2\tilde{M}\left(  \tilde{M}+\left(  \tilde{M}^{2}-\tilde{q}%
^{2}-\frac{\tilde{J}^{2}}{\tilde{M}^{2}}\right)  ^{1/2}\right)  -\tilde{q}^{2}
\label{AMJq}%
\end{equation}
Differentiation of this equation results in
\begin{equation}
d\tilde{M}=\frac{\tilde{\kappa}}{8\pi}dA+\tilde{\Omega}d\tilde{J}+\tilde{\Phi
}d\tilde{q}%
\end{equation}
where
\begin{equation}
\tilde{\Omega}=\frac{\partial\tilde{M}}{\partial\tilde{J}}=4\pi\frac{\tilde
{J}}{\tilde{M}A},\;\;\tilde{\Phi}=\frac{\partial\tilde{M}}{\partial\tilde{q}%
}=\frac{2\pi\tilde{q}^{3}}{\tilde{M}A}+\frac{\tilde{q}}{2\tilde{M}}%
\end{equation}
and
\begin{equation}
\tilde{\kappa}=\frac{4\pi}{A}\left(  \tilde{M}^{2}-\tilde{q}^{2}-\frac
{\tilde{J}^{2}}{\tilde{M}^{2}}\right)  ^{1/2} \label{SurfGrav}%
\end{equation}
represents geometric surface gravity -- free fall acceleration evaluated at
the event horizon and rescaled by so-called red shift to produce finite
values. The red shift is responsible for slowing down time in very strong
gravitational fields. The conventional form of the same differential is
\begin{equation}
dE=\frac{c^{2}}{8\pi\gamma}\kappa dA+\Omega dJ+\Phi dq
\end{equation}
where
\begin{equation}
\Omega=c\tilde{\Omega},\;\;\Phi=\left(  \frac{c^{4}}{4\pi\gamma\varepsilon
_{0}}\right)  ^{1/2}\tilde{\Phi}%
\end{equation}

All equations simplify in case of \textbf{Schwarzschild black holes
}\cite{Davies1978}, which do not have rotation or electric charge: $J=0$ and
$q=0$. A Schwarzschild black hole is spherically symmetric; its properties are
determined by one parameter --- the mass $M$:
\begin{equation}
A=16\pi\tilde{M}^{2}=16\pi\frac{\gamma^{2}}{c^{4}}M^{2},\;\;\tilde{\kappa
}=\frac{1}{4\tilde{M}},\;\;\kappa=\frac{c^{4}}{4\gamma}\frac{1}{M}%
\end{equation}

%
%

\begin{center}

\begin{figure}[h!]
\includegraphics[width=14cm]{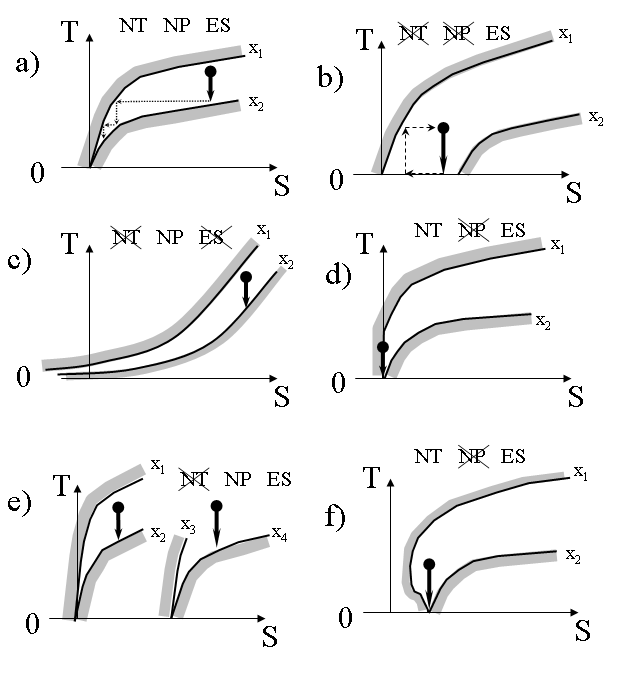}
\caption{Entropy behavior near absolute zero. The lines of constant $x$ bounding isentropic expansion are shown.
Violations of the Nernst theorem (NT), Nernst principle (NP) and Einstein statement (ES) are indicated above each figure.
The dotted arrows in figure (a) show the infinite number of steps needed to reach absolute zero by a sequence of isentropic and isothermal processes.  
The dashed arrows in figure (b) demonstrate the Carnot-Nernst cycle.}
\label{fig1}
\end{figure}

\begin{figure}[tbp]
\includegraphics[width=14cm]{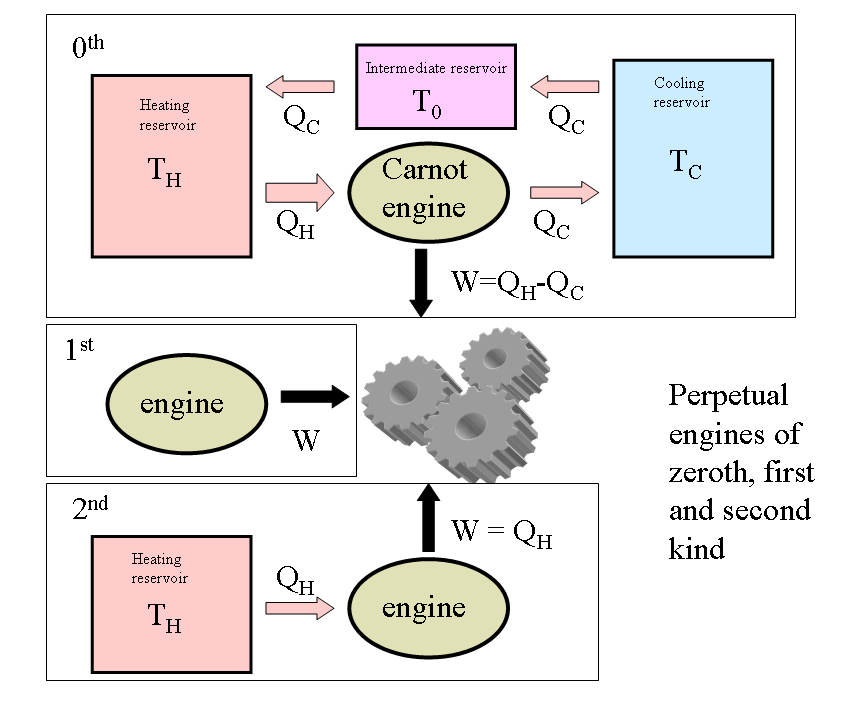}
\caption{Perpetual motion engines of the zeroth, first and second kinds}
\label{fig2}
\end{figure}

\begin{figure}[tbp]
\includegraphics[width=14cm]{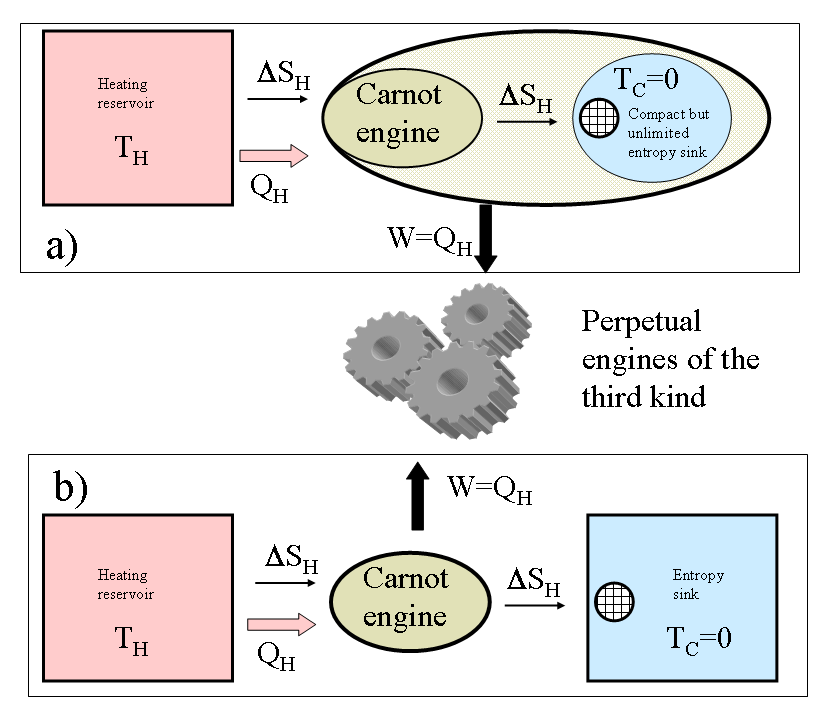}
\caption{Perpetual motion engines of the third kind: a) violating the Einstein statement and b) violating the Nernst principle }
\label{fig3}
\end{figure}

\begin{figure}[tbp]
\includegraphics[width=14cm]{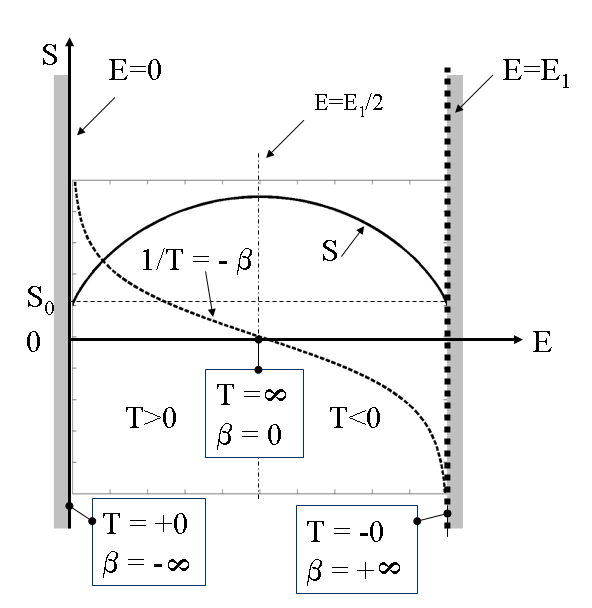}
\caption{Negative temperatures in thermodynamics}
\label{fig4}
\end{figure}

\begin{figure}[tbp]
\includegraphics[width=12cm]{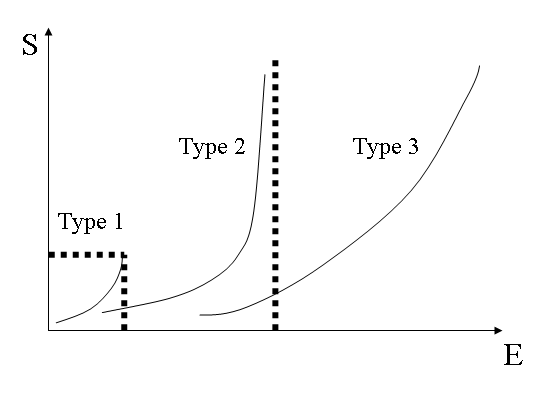}
\caption{Three types of thermodynamic black holes}
\label{fig5}
\end{figure}

\end{center}
\end{document}